\documentclass[useAMS, usegraphicx, usenatbib]{mn2e}

\usepackage[fleqn]{amsmath} 

\title[Radial orbit instability in dwarf dark matter haloes]
{Radial orbit instability in dwarf dark matter haloes}

\author[G. Gajda et al.]
    {Grzegorz Gajda,$^{1,2}$ Ewa L. {\L}okas$^{2}$ and Rados{\l}aw Wojtak$^{3,4}$
    \\
    \\
    $^1$Astronomical Observatory of the Jagiellonian University, Orla 171, 30-244 Cracow, Poland \\
    $^2$Nicolaus Copernicus Astronomical Center, Bartycka 18, 00-716 Warsaw, Poland  \\
    $^3$Dark Cosmology Centre, Niels Bohr Institute, University of Copenhagen, Juliane Maries Vej 30,
    DK-2100 Copenhagen, Denmark \\
    $^4$Kavli Institute for Particle Astrophysics and Cosmology, Stanford University,
    SLAC National Accelerator Laboratory, Menlo Park,\\ CA 94025, USA}

\begin{document}
\maketitle

\begin{abstract}
Using $N$-body simulations we study the phenomenon of radial orbit instability occurring in dark matter haloes of the
size of a dwarf galaxy.
We carried out simulations of seven spherical models, with the same standard NFW density profile but
different anisotropy profiles of particle orbits.
Four of them underwent instability: two with a constant positive anisotropy, one with an anisotropic
core and an isotropic envelope and one with a very small isotropic core and an anisotropic envelope.
Haloes affected by the instability become approximately axisymmetric and prolate,
with the profile of the shortest-to-longest axis ratio increasing with radius. The lower limit for the
central value of this axis ratio is 0.3 for an NFW halo.
The density profiles of the haloes did not change significantly, whereas the velocity distributions became axisymmetric.
The angular momentum modulus rose due to large-amplitude oscillations of its components
perpendicular to the symmetry axis of the halo.
We also studied orbits of individual particles assigning them to classical orbit families in triaxial potentials.
We find that the membership of a given particle in a family depends on its initial angular momentum modulus
and its components along the principal axes of matter distribution.
\end{abstract}


\section{Introduction}

Simulations of structure formation in the Universe with dark matter only lead us to believe that virialized structures,
such as dark matter haloes, possess universal properties that depend only weakly on mass. One such property, most often
used, is the spherically averaged density profile that can be approximated by the well-known NFW \citep*{nfw95} formula.

However, dark matter haloes are definitely not spherical and their shapes can be approximated by triaxial ellipsoids,
with principal axes labelled $a$, $b$, $c$, so that $a>b>c$.
\citet{bailin_steinmetz05} reported that on average $c/a=0.6\pm0.1$ and $b/a=0.75\pm0.15$.
In addition, the longest axis is usually aligned with the closest filament in the neighbouring structure.
\citet{vera-ciro11} showed that the profiles of axis ratios are increasing functions of radius,
i.e. their outer parts are more spherical than the inner ones.
The inner regions are usually elongated, whereas the outer ones are triaxial or even oblate.

Another quantity often used to describe the properties of dark matter haloes is the anisotropy parameter
$\beta = 1 - (\sigma_\theta^2 + \sigma_\phi^2)/(2 \sigma_r^2)=1-(\overline{v_\theta^2} + \overline{v_\phi^2})/(2\overline{v_r^2})$ \citep{binney_termaine08} that
characterizes the type of orbits dominating in the halo. In general, this parameter is also found
to be a function of radius and may depend on mass and environment.
For example, \citet{wojtak08} found that the inner parts of simulated cluster-size haloes are characterized by
$\beta(r=0)=0.05\pm0.05$ while the outer parts are usually radially biased with
$\beta(r_\rmn{v})=0.35\pm0.25$ (where $r_\rmn{v}$ is the virial radius).
Although the nearly zero value of the anisotropy parameter in the central parts of simulated haloes
is commonly interpreted as a signature of an isotropic velocity distribution, it results from spherical
averaging of an axisymmetric and highly anisotropic velocity distribution \citep{wojtak13}.

The origin of these properties and the relation between the density distribution, shape and orbital properties of
dark matter haloes is still under investigation. One phenomenon that may contribute to the evolution of these properties
is the one of orbit instability.
\citet{antonov73} proved that a system consisting entirely of matter streams moving radially is unstable
against perturbations of potential.
Independently, \citet{henon73} showed that a system with a radially anisotropic distribution function is unstable.
The existence of this kind of instability in an $N$-body simulation was demonstrated
for the first time by \citet{polyachenko81}.
He showed that the shape of an initially spherical system, whose particles have only radial component of velocity
(i.e. $\beta=1$), changes dramatically in the course of the simulation.
This type of instability is referred to as the Radial Orbit Instability (ROI).

\citet{polyachenko_shukhman81} analyzed a finite-size system, whose $\beta(r)$ profile rose from zero
in the centre to a given value at the boundary of the system.
They decomposed the density and potential into a complete set of functions with the angular part
described by spherical harmonics $Y_{lm}(\theta, \phi)$ and searched for exponentially growing eigenmodes.
The system turned out to be unstable to the $l=2$ mode.
Moreover, they identified a boundary that separates a system vulnerable to ROI.
They introduced a global parameter $Q_\rmn{global}=2T_{r}/T_{t}$, which for a system without streaming motions of particles amounts to $(1-\overline{\beta})^{-1}$, where $\overline{\beta}=1-T_{t}/(2T_{r})$ is a global anisotropy parameter.
The system studied by \citet{polyachenko_shukhman81} turned out to be stable for $Q_\rmn{global}\la 1.6$, i.e. $\overline{\beta}\la 0.375$.

\citet{merritt_aguilar85} analyzed a suite of systems that all had the same density profile but
various $\beta(r)$ profiles.
They used two families of models: one with $\beta(r)=$ const and one with the Osipkov-Merritt \citep{osipkov79, merritt85} anisotropy profile:
$\beta(r)=(r/r_\rmn{a})^2/[1+(r/r_\rmn{a})^2]$, where $r_\rmn{a}$ is the so-called anisotropy radius.
For constant $\beta$ they found that ROI sets on for $Q_\rmn{global}>1.4$ ($\overline{\beta}>0.29$).
In the case of the second family transition was found for $r_\rmn{a}\approx 0.3r_\rmn{h}$,
where $r_\rmn{h}$ is the half-mass radius. For such a model $Q_\rmn{global}=2.3$ ($\overline{\beta}=0.57$).
Therefore, a global value of $Q_\rmn{global}$ is not a good indicator of the possibility of the onset of ROI.
\citet{merritt_aguilar85} additionally found that the density profile averaged in spherical shells does not change
significantly due to ROI, although the final elongations were $c/a\approx 0.4-0.5$.
Most of their models were axisymmetric, only the one composed only of radial orbits was triaxial, with $c/b=0.7$.

The stability of anisotropic spherical systems against ROI was later investigated also
by means of the adiabatic deformation method \citep{may_binney86}, $N$-body simulations
\citep*{barnes86, merritt87, dejonghe_merritt88, meza_zamorano97}
and the linear stability analysis
\citep{saha91, weinberg91, bertin94}.

ROI may be connected to the form of the density profiles of cosmological haloes.
\citet*{huss_jain_steinmetz99} showed that for a cold collapse occurring in cosmological setting,
if tangential components of gravity were artificially turned off, a halo with the density profile
$\rho\propto r^{-2}$ would form, in contrast with normal conditions when a halo with an NFW profile forms.
In similar simulations \citet*{macmilan_widrow_hendriksen06} noted that the haloes formed were elongated,
with $\beta(r)$ raising from zero in the centre to unity at the virial radius.
One more factor that affects the final shape of the haloes formed from a collapse is the initial virial ratio \citep*[e.g.][]{barnes09}.
\citet{trenti_bertin06} indicated that an initially isotropic core suppresses the onset of ROI during cold collapse.
An additional stabilizing factor is the presence of a black hole in the centre of the system.
It may entirely suppress instability \citep{palmer_papaloizou88} or even drive partial reduction of
elongation \citep{buyle07}.

\citet*{antonini09} found a counterpart of ROI in a system that was already triaxial.
Initially, their system had axis ratios $c/a=0.7$ and $b/a=0.9$.
During the simulation, the shape has changed to less triaxial and finally $c/a\approx b/a \approx 0.65$.
Moreover, reduction of the number of box orbits led to the suppression of instability.

Understanding ROI may be crucial for at least two reasons.
First, it allows us to place constraints on dynamical models constructed to describe real galaxies.
Second, it seems that the onset of ROI may be of importance in the process of the formation of dark matter haloes,
which surround individual galaxies and whole clusters.
In this work we study the impact of ROI on systems having initially the same NFW density profile,
but different anisotropy  profiles. Our study is inspired by recent developments in the understanding of
anisotropy profiles of gravitationally bound systems forming in cosmological context.
We focus on dark matter
haloes of mass and other properties characteristic of present-day dwarf galaxies. This choice is motivated
by the fact that the dynamical time of the systems is relatively short and therefore their evolution may be
significant over cosmic time. Such systems are presumably old, therefore there was likely enough time for
ROI to affect them. Formation of non-spherical dwarf haloes is also important for testing the reliability
of different methods of estimating mass and density profiles of these haloes as such models can be used to generate
mock data samples.

The paper is organized as follows. In section \ref{sec_simulations} we describe the initial conditions used in
our simulations and the set-up of the $N$-body code.
In section \ref{sec_mass_distribution} we show how the distribution of matter in the haloes changes due to the
occurrence of ROI depending of the different anisotropy profiles assumed.
In section \ref{sec_kinematics} we characterize the influence of ROI on halo kinematics, while in section
\ref{sec_individual_orbits} we focus on orbits of individual particles.
The discussion follows in section \ref{sec_discussion} and in section \ref{sec_summary} we summarize our results.

\section{The Simulations}
\label{sec_simulations}

The initial conditions of our simulations were generated as numerical realizations of dark matter haloes with the
same NFW profile and different profiles of the orbital anisotropy.
Since the standard NFW profile does not have a finite mass, as the radial density profile we adopted the NFW
profile with a cut-off at the virial radius $r_{\rmn{v}}$:
\begin{equation}
  \rho(r)=
    \left\{
      \begin{array}{l}
	\frac{\rho_{0}}
	{
	  (r/r_\rmn{s}) \left(1+r/r_\rmn{s}\right)^2
	}
	\quad r<r_{\rmn{v}}
	\\
	\frac{N}
	{
	  (r/r_\rmn{c}) \left(1+r/r_\rmn{c}\right)^5 } \quad r>r_{\rmn{v}
	}
      \end{array}
    \right.
  .
  \label{eq_nfwcut}
\end{equation}
Thus, below the virial radius our haloes have the standard NFW density distribution with the characteristic density
$\rho_0$ and the scale radius $r_\rmn{s}$. Outside the virial radius the density profile is steeper and
the constants $N$ and $r_\rmn{c}$ are determined by the condition that the functions $\rho(r)$ and
$\rmn{d log} \rho/\rmn{d log} r$ are continuous near $r_\rmn{v}$.
We employed haloes with the virial mass (the mass inside the virial radius) $M_{\rmn{v}}=10^9 \rmn{M}_{\sun}$,
the virial radius $25.8$ kpc and the
concentration parameter $c = r_\rmn{v}/r_\rmn{s}=20$, resulting in a scale radius
$r_{\rmn{s}}=1.29$ kpc.
Although these choices make the results directly applicable to
galaxies of the dwarf mass scale, they can be rescaled to more massive objects. The appropriate
scaling formulae for the NFW density profile are given in section 2.1 of \citet{wojtak08}.

We adopted a model of the distribution function proposed by \citet{wojtak08}, namely
\begin{equation}
	f(E,L)=f_E(E)f_L(L),
\end{equation}
with its angular momentum part given by the formula
\begin{equation}
	f_L(L)=L^{-2\beta_0}\left(1+\frac{L^2}{2L_0^2}\right)^{\beta_0 - \beta_{\infty}},
\label{eq_df_lpart}
\end{equation}
where $\beta_0$ and $\beta_{\infty}$ are anisotropies of the system in the centre and at infinity, respectively.
The angular momentum constant $L_0$ is related to the characteristic radius $r_\rmn{t}$ defined as
$\beta(r_\rmn{t})=(\beta_0+\beta_{\infty})/2$ that describes the scale of transition from $\beta_0$ to
$\beta_{\infty}$ in the $\beta(r)$ profile.
Such a profile of anisotropy has three free parameters and is sufficiently flexible to adequately describe
the variability and diversity of $\beta(r)$ profiles found in majority of cosmological haloes.
It is also possible to choose a set of values $(\beta_0, \beta_\infty, L_0)$ that corresponds to a
mean anisotropy profile of a halo \citep{wojtak08}.
The assumed density profile uniquely determines the energy part of the distribution function, $f_E(E)$.
The details of its construction are given in \citet{wojtak08}.

\begin{figure}
  \centering
  \includegraphics[width=\columnwidth]{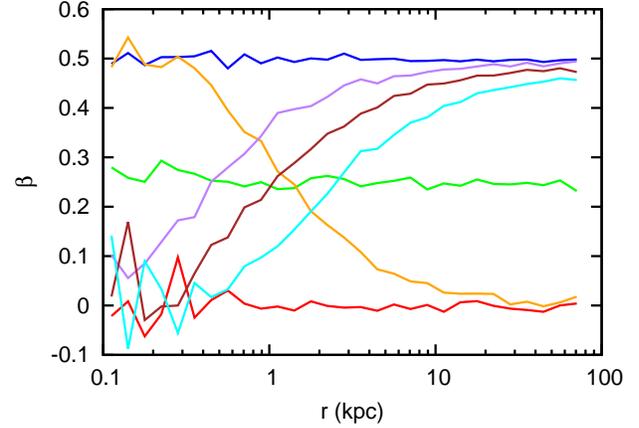}
  \caption{
    The initial anisotropy profiles of simulations C1-C3, D and I1-I3.
  }
  \label{fig_initialanisotropy}
\end{figure}

We generated seven different numerical realizations of dark matter haloes with different initial orbital
anisotropy $\beta$ profiles. The basic properties of the models are listed in Table \ref{tab_models}.
The first three models in the Table, C1-C3, had constant anisotropy parameters: $0$, $0.25$ and $0.5$, respectively.
In the other four the anisotropy varied with radius.
The peculiar model D had an anisotropy profile decreasing with radius from $\beta_0=0.5$ to $\beta_\infty=0$.
The central anisotropy parameter in this model (and model C3) takes the maximum value permitted by the condition of
non-negative phase-space density for a spherical NFW halo (see \citealt*{an_evans06}).
Models I1-I3 had the anisotropy profile increasing with radius, with the same central and asymptotic anisotropy,
namely $\beta_0=0$, $\beta_\infty=0.5$, but the value of the transition radius $r_\rmn{t}$ was varied from
0.5 to 2 times the scale radius $r_\rmn{s}$.
Such a dependence on radius was chosen in order to mimic the anisotropy profiles obtained in cosmological
simulations (see e.g \citealt{wojtak08}).
The last column of Table \ref{tab_models} gives the colour with which the results for a given orbit will be
shown throughout the paper.
All the anisotropy profiles considered here are plotted in Figure~\ref{fig_initialanisotropy}.

\begin{table}
\centering
\caption{Anisotropy models of simulated haloes.}
\label{tab_models}
\begin{tabular}{cllll}
Simulation & $\beta_0$ & $\beta_{\infty}$ & $r_\rmn{t}/r_\rmn{s}$ & Colour \\
\hline
C1  & 0    & 0    & --   &  red   \\
C2  & 0.25 & 0.25 & --   &  green     \\
C3  & 0.5  & 0.5  & --   &  blue    \\
D   & 0.5  & 0    & 1.0  &  orange  \\
I1  & 0    & 0.5  & 0.5  &  purple  \\
I2  & 0    & 0.5  & 1.0  &  brown   \\
I3  & 0    & 0.5  & 2.0  &  cyan   \\
\hline
\end{tabular}
\end{table}

The simulations were carried out using the parallel, tree $N$-body code \textsc{Gadget-2} \citep{springel05}.
Particle mesh lattice of size $128^3$ was used.
The gravitational force softening was set to $0.06$ kpc, so we could reliably resolve physics on the scales
larger than about $0.2$ kpc.
Evolution of $10^6$ particles in each halo was traced for 10 Gyr and we saved their positions and velocities
every $0.05$ Gyr.

\section{Mass distribution}
\label{sec_mass_distribution}

In order to accurately measure the shape of a halo, it is necessary to determine its centre of mass.
We do this by calculating the centre of mass of particles inside a sphere of decreasing radius
until convergence is reached.
The shape of each halo was determined iteratively using the algorithm described by \citet{zemp11}
with a slight modification.
In each step the space was divided into ellipsoidal shells which were allowed to have different axis ratios
and principal axes.
Each shell $S_{i}$ was defined, in a coordinate system aligned with its principal axes, as a set of points
\begin{equation}
\begin{split}
S_i\colon=\left\{(x,y,z)\colon
\log\left[
\sqrt[3]{p_\rmn{ i} q_{i}}\sqrt{ x^2 + \left(\frac{y}{p_{i}}\right)^2 + \left(\frac{z}{q_{i}}\right)^2}
\right]\right. \\
\left. \in \left(\log r_{\rmn{ell}, i}-\log k, \log r_{\rmn{ell}, i}+\log k\right)
\vphantom{\left[\sqrt{\left(\frac{y}{p_{i}}\right)^2}\right]}
\right\},
\end{split}
\end{equation}
where $\log r_{\rmn{ell}, i}$ were equally spaced in the logarithmic space and $k$ was the spacing.
The shells had the following principal axes
\begin{align}
a_{\rm i}&=r_{\rmn{ell}, i}\left(\frac{1}{p_{i} q_{i}}\right)^{1/3}, \\
b_{i}&=r_{\rmn{ell}, i}\left(\frac{p_{i}^2}{q_{i}}\right)^{1/3}, \\
c_{i}&=r_{\rmn{ell}, i}\left(\frac{q_{i}^2}{p_{i}}\right)^{1/3}.
\end{align}
This choice ensures that $b_{i}/a_{i}=p_{i}$ and $c_{i}/a_{i}=q_{i}$.
The volume of the shell $V_i\propto a_i b_i c_i=r_{\rmn{ell},i}^3$, thus we can adopt $r_{\rmn{ell}, i}$
as a mean radius of the $i$-th shell. For each shell the shape tensor
\begin{equation}
S_{jk}=\sum_i (r_i)_j (r_i)_k
\end{equation}
was calculated, where $(r_i)_k$ denotes the \textit{k}-th component of the \textit{i}-th particle position
vector and summation runs over all particles in the shell.
Square roots of eigenvalue ratios correspond to the axis ratios $p$ and $q$.
The direction of the eigenvector of the largest eigenvalue corresponds to the direction of the
longest axis of mass distribution, and so forth.
Afterwards, particles were reassigned to shells and the iterative procedure continued until convergence.

\begin{figure}
  \centering
  \includegraphics[height=0.9\textheight]{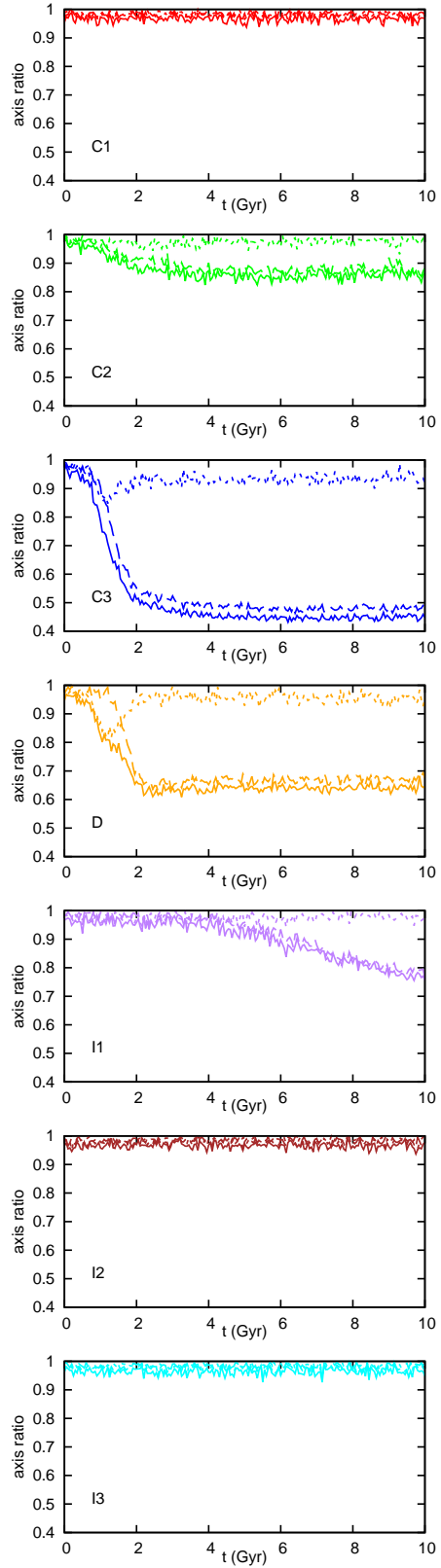}
  \caption{Evolution of the axis ratios measured at $r=r_\rmn{s}=1.29$ kpc. Each panel shows results for a different
    simulation, as marked in the lower left corner.
    Solid lines: shortest to longest ($c/a\equiv q$), dashed lines: intermediate to longest ($b/a\equiv p$),
    dotted lines: shortest to intermediate ($c/b \equiv q/p$).}
  \label{fig_ell_axes_time}
\end{figure}

In Figure~\ref{fig_ell_axes_time} we show the time dependence of different axis ratios measured at
$r=r_\rmn{s}=1.29$ kpc.
The shape of the halo in simulation C1 was not altered, whereas the other two models with constant $\beta(r)$
underwent ROI.
In both cases the evolution begun shortly after the start of the simulation and the shape settled relatively quickly.
As $c/b>0.9$, we conclude that both haloes are approximately axisymmetric and prolate.
Not surprisingly, ROI manifested itself more strongly in model C3 with the largest value of constant anisotropy:
the halo is more elongated and its evolution proceeded more rapidly.

In the case of model D (with decreasing $\beta(r)$) the situation was similar to model C3: the evolution begun
shortly after the start of the simulation, lasted for approximately 1 Gyr and the final shape was slightly
less elongated. One major difference is that initially, until $t\approx 1$ Gyr, the halo was axisymmetric,
however it was not prolate, but oblate ($b/a\approx 1$).

Among the haloes with the increasing profile of anisotropy, only the one with the smallest transition scale
$r_\rmn{t}$ underwent ROI.
Moreover, the onset of the instability was quite slow in this case, the halo remained stable for about 5 Gyr
and did not finish its evolution before the end of the simulation.
As in the previous cases, the final shape is nearly axisymmetric and prolate.

We have also investigated the time evolution of axis ratios at other radii.
It turned out that for models C2, C3 and D the instability manifested itself first in the very centres
of the haloes and then the further from the centre, the later the shape change occurred and the slower was the
rate of this change.
In the case of model D the shape was oblate at first also at other radii and then evolved into prolate.
The evolution of halo I1 was different.
At first, the shape begun to change at radii between $1.5$ and $3.5$ kpc and the central part underwent evolution
only later.

\begin{figure*}
  \centering
  \hfill
  \includegraphics[height=0.85\textheight]{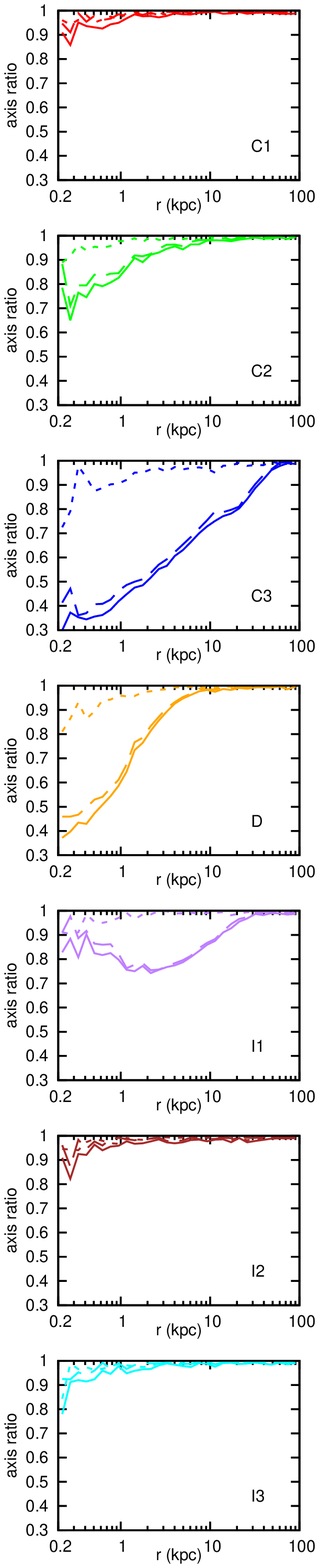}
  \hfill
  \hfill
  \includegraphics[height=0.85\textheight]{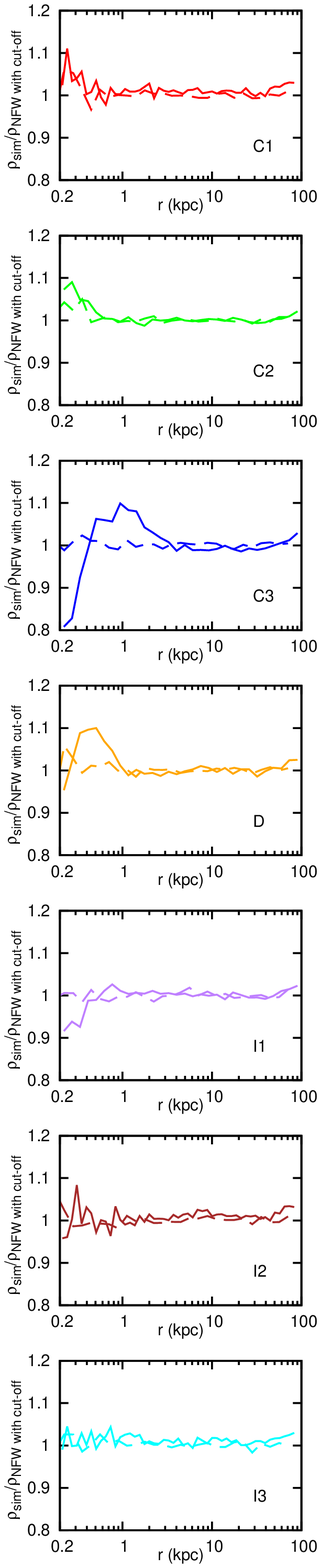}
  \hfill
  \hfill
  \includegraphics[height=0.85\textheight]{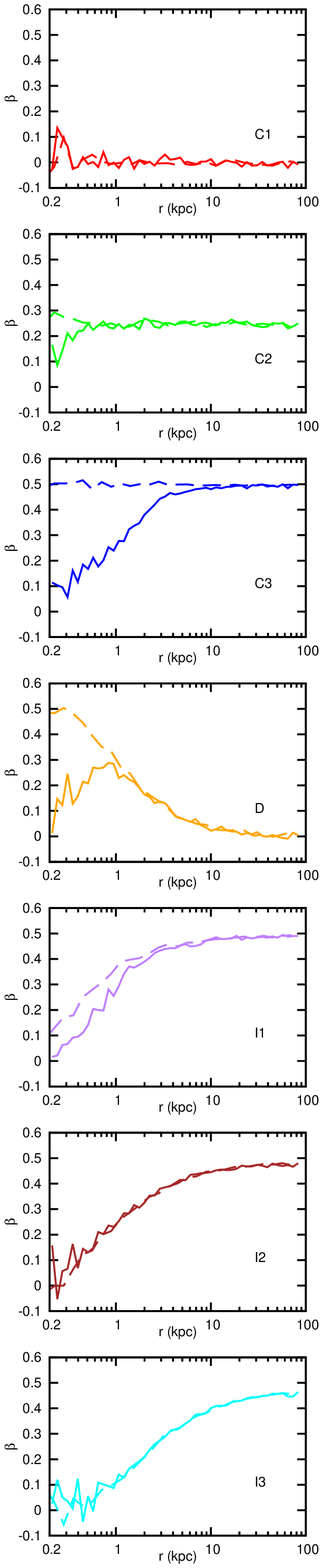}
  \hspace*{\fill}
  \caption[]{
    \emph{Left column}: Final profiles of the axis ratios for all models.
    Solid lines: shortest to longest ($c/a\equiv q$), dashed lines: intermediate to longest ($b/a\equiv p$),
    dotted lines: shortest to intermediate ($c/b \equiv q/p$).
    \emph{Middle column}: Ratio of the density profile of the simulated halo to the density profile given by the
    NFW with a cut-off. Dashed lines: the density profile at the beginning of the simulation calculated in
    spherical shells. Solid lines: the density profile at the end of the simulation calculated in ellipsoidal
    (models C2, C3,I1, D) or spherical (models C1, I2, I3) shells.
    \emph{Right column}: Profiles of the anistropy parameter $\beta$: initial (dashed) and final (solid).
  }
  \label{fig_axes_dens_beta}
\end{figure*}

In the left column of Figure~\ref{fig_axes_dens_beta} we plot the final axis ratios of all haloes as a function of
radius.
It may seem that the central parts of models C1, I2 and I3 are elongated, having $c/a\approx 0.9$.
However, it is not a physical feature, but a numerical one.
In the central shells there were so few particles that the algorithm converged slowly and was not very stable.
At the end of the simulation the shapes of haloes C2 and D were stable everywhere.
Halo C3 was still evolving beyond $r\approx 20$ kpc and halo I1 did not finish its evolution anywhere.

It is interesting to check whether the principal axes of different shells are aligned and whether
this alignment evolves in time.
It turns out that for all simulations the situation is similar.
The direction of the longest axis does not depend on radius and is also very stable in time.
If for a given shell the shortest-to-longest axis ratio drops below $\sim 0.97$, the longest axis has
the same direction as in the whole halo.
The intermediate and shortest axis are less stable and wobble by up to $20\degr$ in the plane
perpendicular to the longest axis.
This is likely a numerical effect due to the small difference between their lengths.

We have verified whether the ellipsoid obtained from the diagonalization of the shape tensor corresponds to the
contours of constant density.
In a spherical shell the density may vary in different directions by more than a factor of 2.
In a given ellipsoidal shell, the maximum differences do not exceed ten per cent.

In the middle column of Figure~\ref{fig_axes_dens_beta} we show the final density profiles of the haloes divided by
the initially assumed density profile from equation \eqref{eq_nfwcut} and compare them to the initial ratio.
Since in the haloes which underwent ROI the density is not constant in spherical shells, for simulations C2, C3, I1
and D the density distribution was computed in ellipsoidal shells.

Despite considerable changes in the halo shapes, the density profiles were not significantly modified,
which means that the density distribution can be described by a generalized triaxial NFW profile.
However, we note that for haloes that underwent ROI the density was decreased a little in the centre so
some outward flow of matter must have occurred in the central parts.
The profiles were modified so that in all those cases the central logarithmic slope is slightly shallower than $-1$,
the value characteristic of the NFW profile.

\section{Kinematics}
\label{sec_kinematics}

Although the orbital anisotropy $\beta$ does not possess a clear physical meaning for non-spherical objects
(see e.g. \citealt*{wojtak13}), we checked how it changed during the simulations.
A comparison of $\beta$ profiles at the beginning and at the end of all simulations is shown in the right
column of Figure~\ref{fig_axes_dens_beta}.
It may seem that in regions where ROI has taken place, the anisotropy gradually decreased making the distribution of orbits more isotropic.

\begin{figure}
  \centering
  \includegraphics[width=\columnwidth]{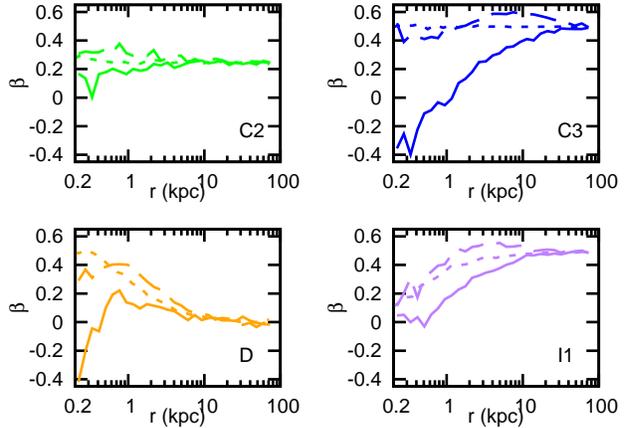}
  \caption{Final directional profiles of the anisotropy parameter $\beta$ for all haloes which
underwent ROI,
calculated along (dashed lines) and perpendicular to (solid lines) their major axes. For comparison
we included also the initial anisotropy profiles (dotted lines).}
\label{fig_beta_nonsph}
\end{figure}

However, it has been noted that the velocity distribution in the centres of haloes has the same symmetry as the matter distribution, i.e. in the axisymmetric haloes the velocity distribution is also axisymmetric \citep{sparre_hansen12, wojtak13}.
In order to examine this, for all of the haloes which have undergone ROI we calculated $\beta$ profiles in cones along and in annuli perpendicular to the major axis of the density distribution.
The results are shown in Figure~\ref{fig_beta_nonsph}, together with the initial $\beta$ profiles.

In both directions the region where the anisotropy parameter changed coincides with the region where the shape of the mass distribution was altered. Further away from the centres of the haloes $\beta$ remained the same as initially.
In the direction perpendicular to the major axis $\beta$ decreased significantly, making the distribution of orbits isotropic (haloes C2, I1) or even tangentially biased in the case of the haloes C3 and D.
On the other hand, $\beta$ profiles along the major axes changed much less.
In the very centres of the haloes they decreased a little and increased in the middle parts of the haloes.

When the $\beta$ parameter is computed in spherical shells, as in the left column of Figure \ref{fig_axes_dens_beta}, the values from different regions are averaged.
Thus it seems that the distribution of orbits is isotropic in the very centres of the haloes and the region where the anisotropy parameter decreased is smaller than the region where the shape was altered.

We have also calculated eigenvalues and eigenvectors of the velocity dispersion tensor.
The vectors are usually consistent with the direction of the principal axes of mass distribution.
The velocity dispersion in the direction of the longest axis is largest and the other two are smaller and almost equal.
Such an alignment of the velocity dispersion tensor is
also seen in cosmological simulations (e.g. \citealt{wojtak13}) and confirms that the velocity distribution is indeed axisymmetric.

\begin{figure}
  \centering
  \includegraphics[width=\columnwidth]{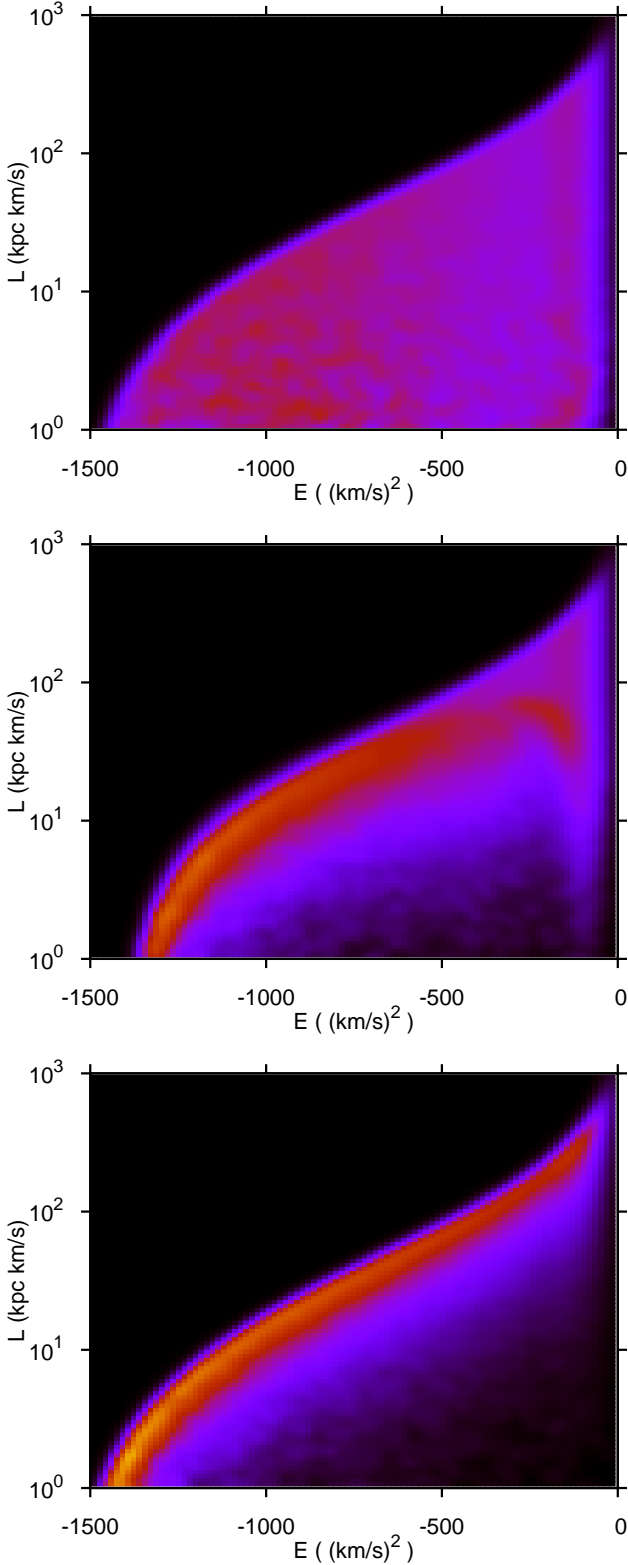}
  \caption{Distribution of particles $N(E, L)$: brighter colours correspond to larger numbers of particles.
  From the top, respectively: the initial distribution for simulation C3, the final distribution for simulation C3
  and the initial distribution for simulation C1 (the isotropic one).}
  \label{fig_distribution_function}
\end{figure}

In the following discussion of the properties of the haloes affected by ROI we concentrated on simulation C3,
as in this case the halo shape changed the most, thus one can expect that other properties also changed significantly.
One interesting question to ask is how the distribution function was affected by the instability.
Unfortunately, it is non-trivial to obtain the distribution function from information about positions and velocities
of particles only.
Note that the distribution function is given by
\begin{equation}
	f(E,L)=\frac{N(E, L)}{g(E, L)},
\end{equation}
where $N(E,L)$ is the number of particles with energy $E$ and angular momentum $L$ and $g(E,L)$ is the
volume of hypersurface in phase space given by $E$ and $L$.
It can be easily calculated only when the analytic form of the gravitational potential is available.
In Figure~\ref{fig_distribution_function} we show instead the $N(E, L)$ itself. Brighter colours correspond to
a larger density of particles.

The top panel was generated from the initial state of simulation C3, whereas the middle one from its final state.
For comparison, in the lower panel we present the same distribution for the isotropic model C1.
The line above which there are no particles indicates the maximum angular momentum at a given energy.
Such a value of $L$ is characteristic of particles on circular orbits.
As a result of the evolution, the distribution for simulation C3 became more similar to the isotropic distribution
due to the increase of the angular momentum of particles in the region where ROI occurred.
In addition, the depth of the potential well decreased as a result of the decrease of density in the centre of the
halo.

\begin{figure}
  \centering
  \includegraphics[width=\columnwidth]{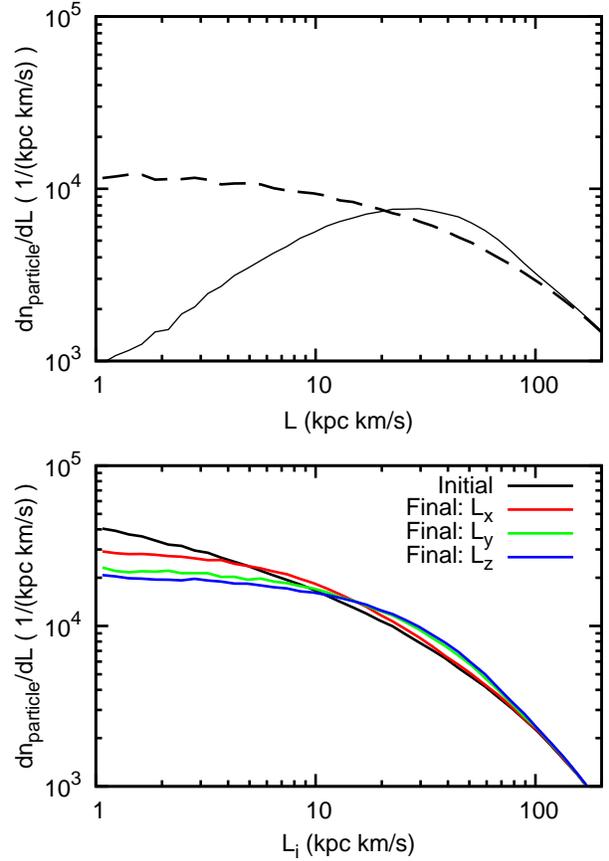}
  \caption[]{
    The distributions per unit of angular momentum for simulation C3 of:
    \emph{Top}:  the angular momentum modulus: initial (dashed line) and final
    (solid line).
    \emph{Bottom}: the initial and final components of the angular momentum.
    The coordinate system is such that $x$, $y$ and $z$ correspond to the major, intermediate and minor axis
    of the halo.
  }
  \label{fig_ang_mom_distribution}
\end{figure}

Since the angular momentum modulus rose distinctly, we analyzed the change of its components.
After undergoing ROI, the halo is no longer spherically symmetric, so we introduced a new coordinate system,
aligned with the principal axes of the matter distribution.
The axes are stable, so for simulation C3 we computed their mean directions and used them to define the
coordinate system so that the $X$ axis is along the longest axis,
$Y$ along the intermediate and $Z$ along the shortest.

In Figure~\ref{fig_ang_mom_distribution} we illustrate the modification of the distribution of
particles with a given angular momentum modulus or its components (per angular momentum unit).
At the end of the simulation, the number of particles with $L\approx 0$ decreased by more than a factor of 10
which means that the angular momentum increased significantly.
The initial distributions of the angular momentum components were the same.
The component $L_x$ increased only slightly during the evolution.
The approximate conservation of this component could be expected, as the shape of the halo is approximately
axisymmetric and so is the potential.
The distributions of $L_y$ and $L_z$ were subject to significant evolution and as a result the
angular momentum has grown.

\section{Orbits of particles}
\label{sec_individual_orbits}

Knowing that for individual particles the angular momentum rose, we may ask how exactly it changed and
how particle orbits were affected.
In a spherical potential orbits are always confined to their initial plane and eventually form a rosette-like shape.
In a more general potential the situation is much more complicated due to the fact that they are usually non-integrable,
with a notable exception of a St{\"a}ckel potentials.
\citet*{dezeuuw85} showed that in a perfect ellipsoid there exist four distinct families of orbits:
the box (B), short-axis tube (S), inner long-axis tube (I) and outer long-axis tube (O).
For the figure depicting the shape of every family we refer the reader to \citet{statler87}
or Figure 3.46 in \citet{binney_termaine08}.
If a given potential has a shape similar to the perfect ellipsoid, those families persist, but also various
resonant (R) orbits may be present, as well as chaotic ones.

\begin{table}
\centering
\caption{Conditions for membership in families of orbits}
\label{tab_orbit_conditions}
\begin{tabular}{cll}
Family & Components of \textbf{\textit{L}} & Outermost positions\\
\hline
B       &$\overline{L}_x, \overline{L}_y, \overline{L}_z\approx 0$& Symmetric: $|x_{\rm max}|\approx|x_{\rm min}|$,\\
& & similarly for $y$, $z$ \\
R       &$\overline{L}_x, \overline{L}_y, \overline{L}_z\approx 0$& Asymmetric\\
S       &${\rm sgn} L_z=$ const & --- \\
I       &${\rm sgn} L_x=$ const & $|x(r_{\rm max})|>|x(r_{\rm min})| $\\
O       &${\rm sgn} L_x=$ const & $|x(r_{\rm max})|<|x(r_{\rm min})| $ \\
\hline
\end{tabular}
\end{table}

In an integrable potential in order to classify an orbit it is sufficient to know its integrals of motion,
as orbit families occupy distinct regions in the space of integrals of motion.
In the simulation we do not know the exact form of the potential, moreover it evolves in time,
thus it is impossible to determine integrals of motion and place any definite conditions on the membership
of an orbit in a given family.
There exists an efficient method for searching for basic frequencies
(e.g. \citealt{carpintero_aguilar98}), but it could not be used for our simulations,
as it requires a much denser sampling of particle positions and velocities. Here, we adopt instead the approach of
\citet{schwarzschild93} who described simple criteria that allow to determine the orbit type for a given particle.
We modified them slightly and present them in Table \ref{tab_orbit_conditions}.

As in the previous section, the coordinate system is such that the $x$, $y$ and $z$ axis are
oriented along, respectively, the major, intermediate and minor axis of the matter distribution.
For the B family it is crucial for all of the mean components of the angular momentum
$\overline{L}_x$, $\overline{L}_y$ and $\overline{L}_z$ to vanish. In addition, this kind of orbit
is symmetric with respect to the principal planes of the coordinate system. This means that the
largest distances from both sides of the $YZ$ plane, i.e. in the positive direction of $x$ axis
($x_\rmn{max}$) and negative ($x_\rmn{min}$), are the same (except for the sign):
$|x_\rmn{max}|\approx|x_\rmn{min}|$. A similar condition holds for the outermost positions in the
$y$ and $z$ direction. The simple resonant orbits from the family R also have negligible
$\overline{L}_x$, $\overline{L}_y$ and $\overline{L}_z$, but they are asymmetric with respect to at
least one plane of the coordinate system.

The three remaining families have a fixed sense of rotation and, consequently, one component of the
angular momentum does not change its sign, thus its average is non-vanishing. For the family S the
non-alternating angular momentum component is $L_z$. The distinction between the two families of
orbits rotating around the $x$-axis (both of which have $\rmn{sgn}L_x=$ const) is given by a
simple feature, which can be easily seen in the relevant figures of \citet{statler87}: the I
orbits are \emph{concave}, while the O orbits are \emph{convex} with respect to the $x$ axis (i.e.
their axis of revolution). Quantifying this feature results in that for family I the maximum
distance from the $x$-axis ($r_\rmn{max}$) and the minimum one ($r_\rmn{min}$) occur, respectively,
at a large and a small distance from the $YZ$ plane. Simply put, $|x(r_\rmn{max})|>|x(r_\rmn{min})|$.
The situation is reversed for the family O.

For the analysis of orbits we again chose simulation C3 because it underwent the most pronounced evolution,
thus we expect the largest differences between orbit types.
We have analyzed orbits of all particles with initial energy in the range $800 \pm 2$ (km
s$^{-1}$)$^2$. There were about 1200 such particles. In the NFW potential the circular orbit of
such energy has the radius of about 1.95 kpc and the period of 0.67 Gyr. Its angular momentum, and
thus the maximum angular momentum for such energy, amounts to $L_\rmn{max}=35.7$ kpc km s$^{-1}$.
At this distance from the centre, the shape ceases to change at about 2.5 Gyr from the beginning of
the simulation, so the adopted conditions from Table \ref{tab_orbit_conditions} apply to a later
time.

Such a choice of the energy is a compromise between two extremes.
Particles further away from the centre of the halo completed too few cycles to reliably determine the families they belong to.
On the other hand, particles from the more central part of the halo have short periods and with positions saved only every $0.05$ Gyr we could not fully trace their motion.

\begin{figure*}
  \centering
  \includegraphics[height=0.225\textheight]{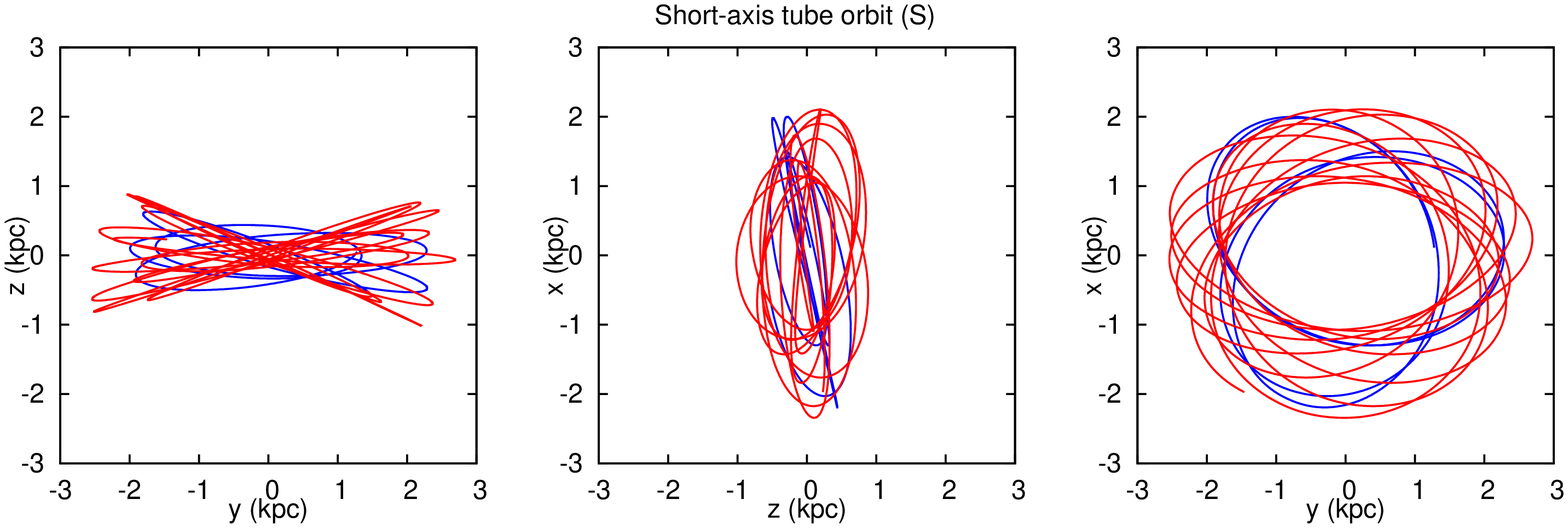} \\
  \includegraphics[height=0.225\textheight]{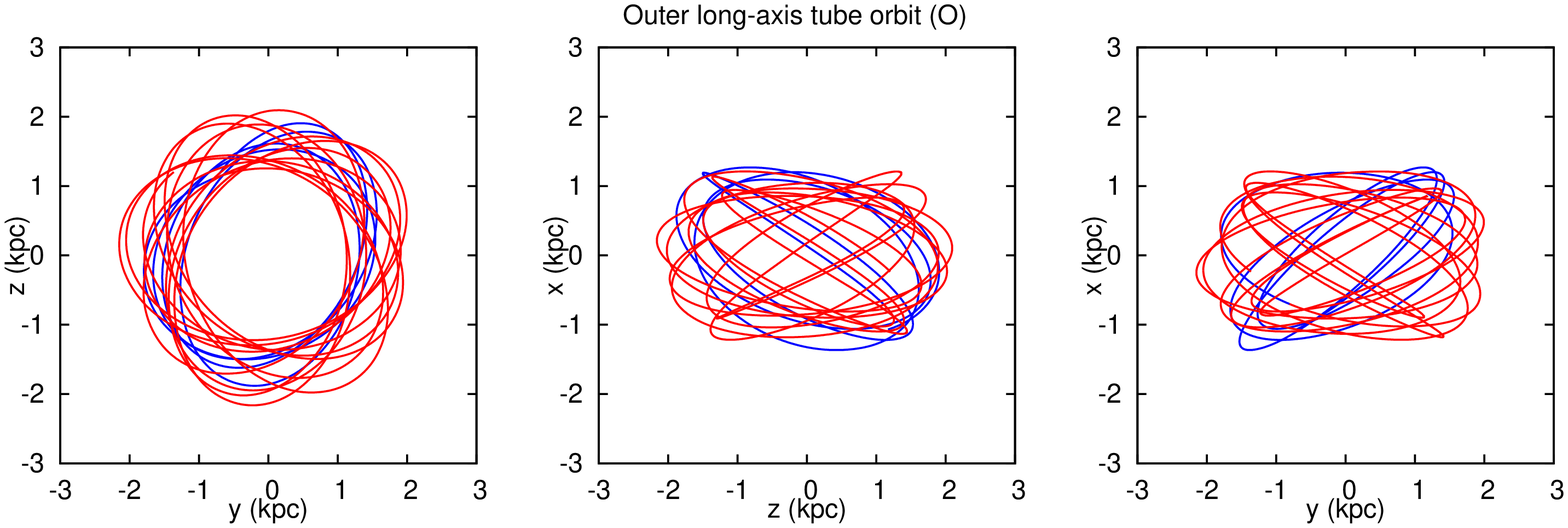} \\
  \includegraphics[height=0.225\textheight]{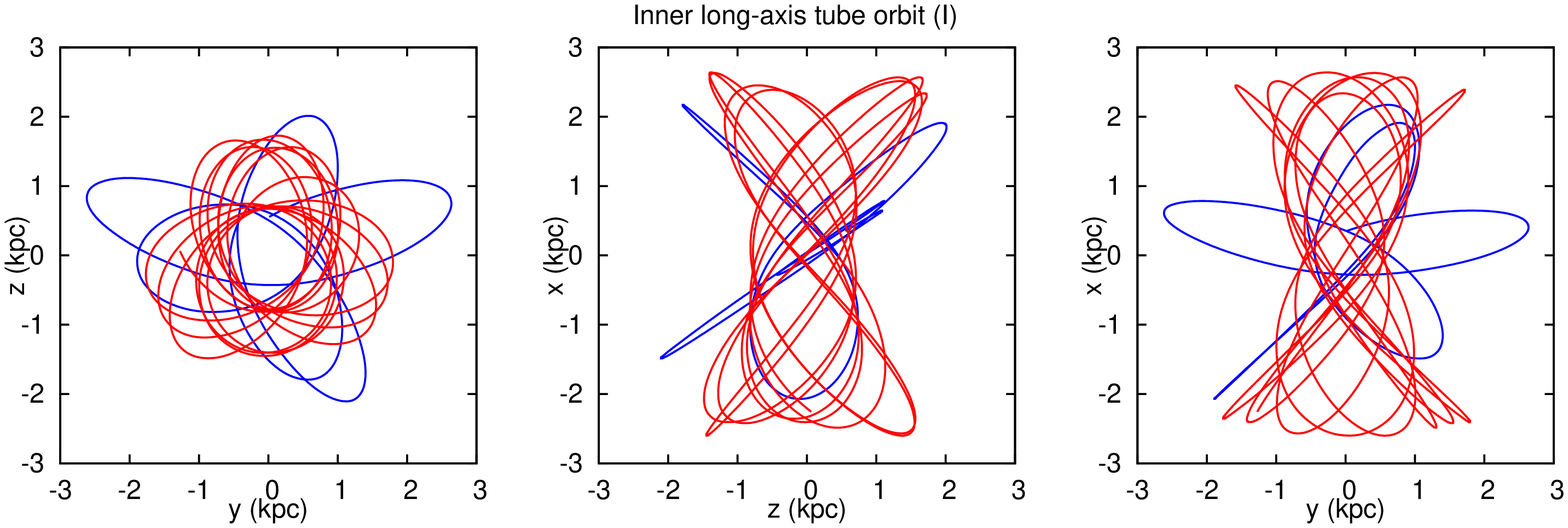} \\
  \includegraphics[height=0.225\textheight]{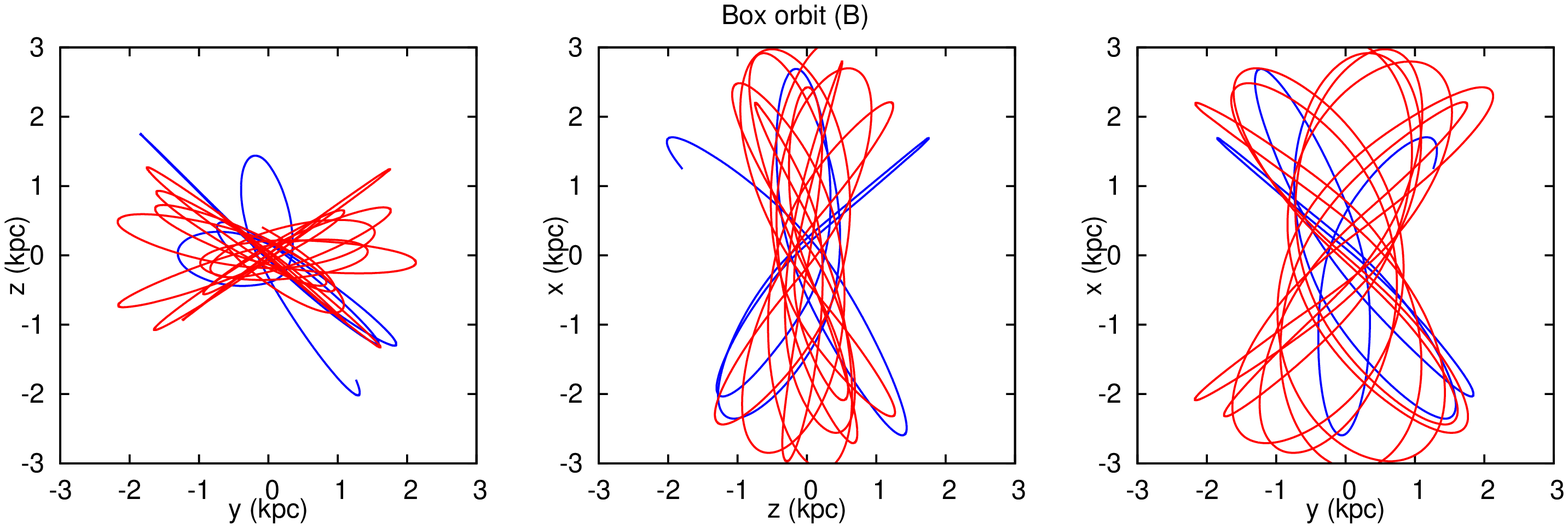}
  \caption{Examples of projections of trajectories for particles belonging to different orbit families
  in simulation C3. The coordinate system is such that $x$, $y$ and $z$ are along the major,
  intermediate and minor axis of the halo.
  The two colours correspond to different periods of evolution: $t<2.5$~Gyr (blue) and $t>2.5$~Gyr (red).}
  \label{fig_orbit_projections}
\end{figure*}

In Figure~\ref{fig_orbit_projections} we show a few examples of particle orbits in projection
onto the principal planes of the coordinate system.
In most cases the change of the orbit after $t \approx 2.5$ Gyr is easily seen.
Apart from stability, a system made up only of O-type particles would be axisymmetric and oblate
(but then we would call this family S).
In the case of family I, it would be prolate and, in an extreme case, slightly triaxial.
The contribution from the box orbits also makes the system elongated, but much more triaxial.

It is worth noting that the extreme position in the $x$-direction of a B-type particle is larger than
in the case of a particle belonging to family I.
It is a result of an I-type particle having a non-zero velocity at a maximum distance
(to conserve $L_x=$const$\neq 0$), whereas a B-particle stops at this point, thus it can move away
more while conserving its energy.
In the cases of both families I and B, a particle spends more time at a greater distance from the $XY$-plane
than in its vicinity. The situation is similar to that in a Keplerian potential when it spends more time
away from the centre of the potential.

\begin{figure}
  \centering
  \includegraphics[height=0.9\textheight]{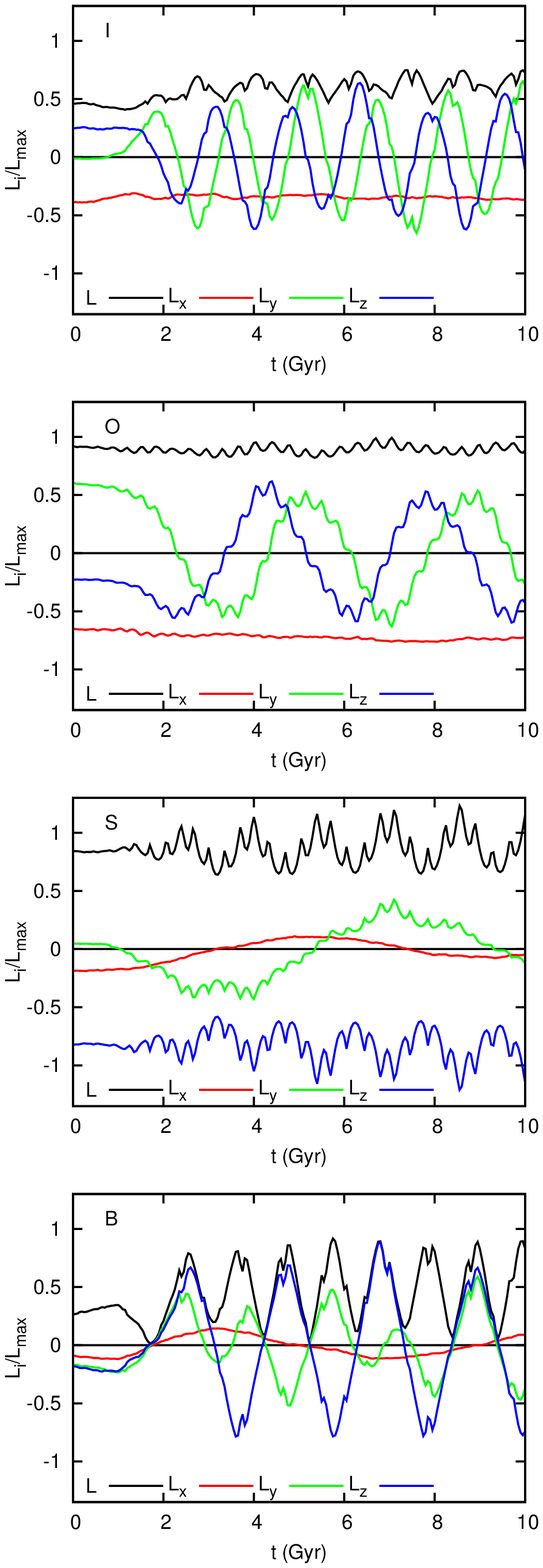}
  \caption{The time dependence of the angular momentum for particles belonging to different orbit families
  in simulation C3. The coordinate system is such that $x$, $y$ and $z$ are along the major,
  intermediate and minor axis of the halo.}
  \label{fig_particle_angular_momentum}
\end{figure}

In Figure~\ref{fig_particle_angular_momentum} we show the time dependence of the
modulus of the angular momentum and its components for the same particles that were used in Figure~\ref{fig_orbit_projections}.
For the family I, the $L_x$ component is approximately conserved, but the other two oscillate in time.
In consequence, the angular momentum modulus also oscillates and is usually larger than its initial value.
The variation of the angular momentum of the O-type particles is similar.
In the case of the S-type orbit, the sign of the $L_z$ component is conserved and the other two oscillate.
For the box and resonant orbits all three components oscillate with large amplitudes and their mean values are zero.
In effect, the mean angular momentum modulus is larger than at the beginning.
The energy of each single particle is approximately constant, however non-periodic changes are present.
This is related to the time-dependence of the potential due to the evolution of the halo shape, hence the energy is
not an integral of motion.

One can raise the question whether we can predict which family a given particle will join based on its initial
position and velocity vectors.
If the families differ between themselves in angular momentum, then such classification will likely depend on
the initial angular momentum vector.

\begin{figure}
  \centering
  \includegraphics[width=\columnwidth]{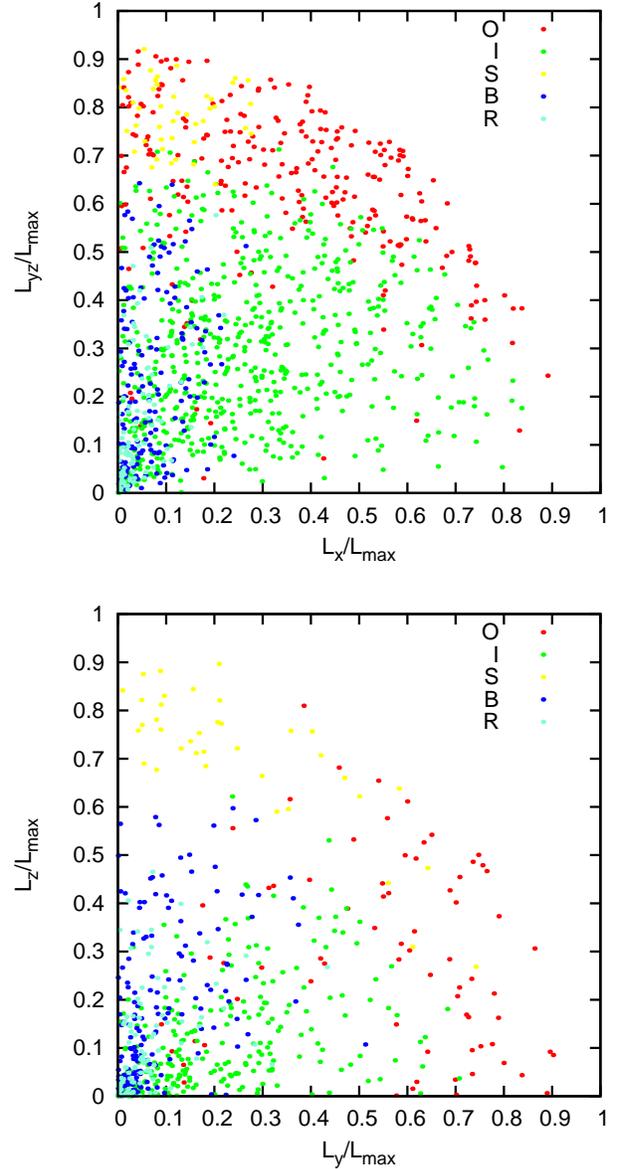}
  \caption[]{
    \emph{Top panel}: Orbit classification, depending on the initial $L_x$ and $L_{yz}=(L_y^2+L_z^2)^{1/2}$.
    \emph{Bottom panel}: Orbit classification for particles which originally had $L_x/L_\rmn{max}<0.2$,
    depending on the initial $L_y$ and $L_z$.
  }
\label{fig_orbit_classification}
\end{figure}

In the top panel of Figure~\ref{fig_orbit_classification} we show to which family a particle belongs,
depending on its initial $L_x$ and $L_{yz}$, i.e. the projection of the angular momentum vector onto the $YZ$ plane.
First, let us consider the region $L_x/L_\rmn{max}\ga 0.3$. Clearly, the families I and O occupy distinct regions.
For small $L_x$ particles belong to family O above $L/L_\rmn{max}\approx 0.7$.
For large $L_x$ this value raises up to $0.8$ or maybe even more.
Interestingly, I orbits with initial $L_x/L_\rmn{max}>0.7$ are actually oblate -- their maximum
distance from the $YZ$ plane is smaller than the maximum distance from the $X$ axis.
Hence all particles with $L/L_\rmn{max}>0.7$ appear to eventually move on an orbit with an oblate shape.
Particles from the family O residing in the uniform I region are erroneously interpreted
resonances. We also note that not all the particles which were marked by our
algorithm as belonging to the family I indeed belong to it, they just have similar shape. It is
due to the fact that our orbit classification is very simple and based on very crude properties
of the angular momentum time dependence and the shape of the orbit of the particle.  It is not
precise enough to grasp the whole variety of resonant and chaotic orbits.

In order to examine the region $L_x/L_\rmn{max}\la 0.2$, in the bottom panel of Figure~\ref{fig_orbit_classification}
we plot all particles fulfilling this condition, in the plane of the initial $L_y$ versus $L_z$.
For $L_{yz}/L_\rmn{max}>0.7$ two families are present: S for $L_z/L_\rmn{max}\ga 0.6$ and O in the opposite case.
For small values of $L_{yz}/L_\rmn{max}$ there is no clear division between orbits from families I, B and R.
The orbits B and R seem to disappear for $L_y/L_\rmn{max}\ga 0.35$.
In the case of family I the larger the ratio $L_z/L_y$ (above unity), the smaller the fraction of particles with
this type of orbit among all. A closer inspection reveals that they usually then have larger $L_x$.

In a spherically symmetric NFW potential, the region which the particle is allowed to occupy is uniquely
determined by the energy and the angular momentum vector.
It has to be noted that the final orbit type does not seem to be always determined by the components of the
angular momentum, but may also depend on other factors, such as the particle trajectory when the shape evolution occurs.

\section{Discussion}
\label{sec_discussion}

\begin{figure}
  \centering
  \includegraphics[width=\columnwidth]{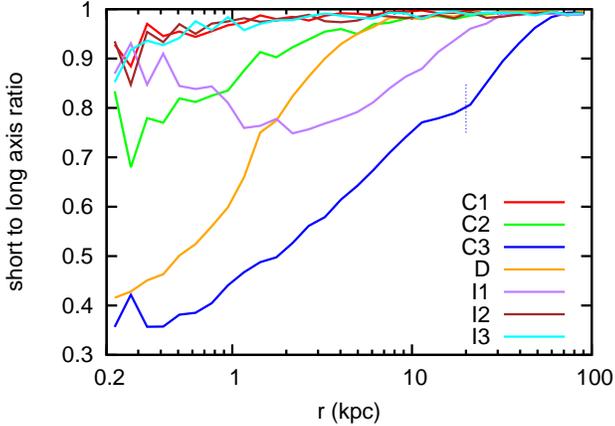}
  \caption{Comparison of the final short-to-long axis ratios, defined as $(c/a+b/a)/2$, for all simulations.
  The vertical dotted line indicates the radius beyond which the evolution of model C3 is not yet completed.
  The evolution of model I1 has not finished anywhere.}
  \label{fig_comparision_axes_ratios}
\end{figure}

All final states of the haloes can be considered biaxial, thus in Figure~\ref{fig_comparision_axes_ratios} we
plot the final short-to-long axis ratios for all models, with the ratio defined as $(c/a+b/a)/2$.
One should notice different areas affected by the instability.
The haloes in simulations C1, I2 and I3 have not undergone ROI. In these cases the axis ratios at all radii are
consistent with unity within uncertainties from shot noise.

First, let us consider models whose initial profile of the anisotropy parameter was constant, i.e. $\beta(r)=$const.
Systems with an isotropic distribution of orbits ($\beta=0$), such as model C1, are stable and ROI does not
develop in them. The more radial the orbits, i.e. the larger is the initial $\beta$ parameter, the greater is the
impact of ROI on the shape of the halo.
The moment when the configuration of the particles begins to change depends to some extent on the initial anisotropy,
whereas the rate of this change depends significantly on $\beta$.
After some time, when the new orbits of particles become stable, the halo reaches its final,
approximately axisymmetric shape, with the axis ratios increasing with radius.

If we introduce the triaxiality parameter $T=(a^2-b^2)/(a^2-c^2)$ we can quantify the degree of triaxiality in the haloes.
The haloes in simulations C2 and C3 have $T>0.9$ in the region subject to instability, hence we can conclude they
are biaxial and prolate.
It may seem that they are more triaxial in the centres, but the determination of the axis ratio in the central
part of the halo is uncertain. Additionally, haloes with a larger initial anisotropy are slightly more triaxial.
In both haloes, at the end of the simulation $\beta(r=0)\approx 0$, however only in the very central part of the halo
the distribution of orbits is more isotropic than initially.
There exists a vast region where $\beta$ is almost equal to its initial value, though the shape of the matter
distribution has altered significantly.

The models C3 and D had the initial central anisotropy $\beta_0=0.5$, but for the latter $\beta(r)$ was
a decreasing function of radius, approaching the isotropic distribution at large radii.
Comparing these simulations we conclude that for model D the rate of the evolution was considerably
slower and affected a smaller part of the halo.
The more isotropic distribution of orbits in the outer part of the halo suppressed the development of the
instability: the radius beyond which the halo remained spherical ($r \approx 8$ kpc) corresponds to the radius
at which the orbits are practically isotropic ($\beta<0.05$).

The central axis ratio in simulation D amounts to about 0.4 and in simulation C3 to a little more than $0.3$.
Taking into account that for the NFW profile the maximum $\beta_0=0.5$ (and that even using an
increasing profile of $\beta(r)$ with $\beta_\infty=1$ would probably not change the situation much compared to
model C3), we can formulate a hypothesis that the minimum central axis ratio that can occur in an NFW halo
as a result of ROI is about 0.3.
Interestingly, it is consistent with the maximum elongation of elliptical galaxies, which are similar to dark matter
haloes in a sense of being supported by random motion of their constituents.
In the case of elliptical galaxies it is argued that the bending instability prevents the formation of
more elongated shapes \citep{merritt99}. It is thus plausible that the occurrence of this instability halts the
development of ROI.

Among haloes which initially had an isotropic distribution of orbits in the centres and anisotropic in the outer parts,
only model I1 underwent ROI. The remaining two models were stable during the 10 Gyr of the simulation.
We cannot rule out that extending the simulation to a longer period would lead to the occurrence of the instability.
Then the situation would be similar to the case studied by \citet{meza_zamorano97} who showed that one of the models
claimed by \citet{merritt_aguilar85} to be stable turns out to be unstable after doubling the simulation time.

\citet*{polyachenko11} argued that for generalized polytropes with a distribution function of the form
$f(E,L)\propto L^{-2\beta} E^{q}$ if $\beta>0$ then a growing eigenmode is always present, but the growth rate
$\gamma$ decreases exponentially with $\beta$, i.e. $\gamma\propto\exp(-\beta_*/\beta)$.
Hence, in the case of a finite simulation time a given system may seem stable.
In our simulations, model I1 appeared stable for a long time before the onset of the instability, so models
I2 and I3 might have turned out to be unstable as well if only the simulation had lasted longer.

However, there is a major difference between systems with a distribution function whose angular momentum
part is given by equation \eqref{eq_df_lpart} with $\beta_0=0$ and generalized polytropes.
In the latter case the limit $L\to 0$ of $f(E,L)$ is not bounded, and according to \citet{palmer_papaloizou87}
and \citet{polyachenko11} it is a sufficient condition for ROI to occur.

In models with $\beta(r=0)>0$ the instability took place first in the very centres of the haloes
and then spread outwards. In this respect, model I1 was significantly different in a sense that in this case the
ROI manifested itself first at a considerable distance from the halo centre.
A closer inspection revealed that the central part of this halo was not subject to instability, despite the fact that
the axis ratios differed from unity.
The principal axes of the shape tensor in the central part are not aligned with the principal axes in the outer part.
Moreover, the value of $\beta$ in the centre began to decrease only after the decrease in the outer part took place.

A similar effect should be present in the case of the instability occurring in Osipkov-Merritt models.
Unfortunately, the radial dependence of axis ratios was rarely examined in the literature.
In the case of \citet{merritt_aguilar85} work, the central parts reached $c/a\approx 0.77$, whereas at the
radius encompassing 60 per cent of the system mass $c/a\approx 0.33$.
\citet{bellovary08} studied the cold collapse of haloes in cosmological setting and for a system with a
large isotropic velocity dispersion obtained a spherical nucleus.
However, they did not point out that at an average distance from the halo centre $c/a\approx0.4-0.5$.
Since the outer part of their halo remained spherical, their $c/a$ profile was qualitatively similar to the
one obtained in our simulation~I1.

The values of $Q_\rmn{global, cr}=(2T_r/T_t)_\rmn{cr}$, above which the system is unstable against
ROI, reported in the literature depend on the model and are widely scattered in the interval
$1.4-2.6$, which translates to $\overline{\beta}=0.29-0.62$ (see a review by \citealt{merritt99}).
Among our models which underwent ROI the one with the smallest $Q_\rmn{global}$ is model D, from
which we can derive an upper limit for the critical value $Q_\rmn{global, cr}<1.14$ for the NFW
density profile. Interestingly, models I2 ($Q_\rmn{global}=1.63$) and I3 ($Q_\rmn{global}=1.47$)
with a higher value of parameter $Q$ seemed to be stable. \citet{nipoti11} suggested that a
parameter $Q_\rmn{half}\la 1.3$, which is defined in the same way as $Q$, but calculated only
within the half-mass radius, might be a good indicator of the stability. However, our system D had
$Q_\rmn{half}=1.2$ and was unstable.
We note that this value of $Q_\rmn{half}$ should not be treated as a new and reliable constraint
on the occurrence of ROI.

Moreover, from Figure~\ref{fig_comparision_axes_ratios} it is obvious that the area subject to ROI
is smaller if the initial distribution of orbits is more isotropic. It seems that there might exist
a critical profile $\beta_\rmn{cr}(r)$ (or, similarly, $Q_\rmn{cr}$), below which we cannot confirm
the occurrence of ROI, because the $c/a$ and $b/a$ profiles would be indistinguishable from
numerical noise (also present in the profiles for simulation C1). It would not mean that we have
found a true critical value (or profile) of anisotropy, but only that we reached the limit of the
resolution of the numerical results. Additionally, \citet{palmer_papaloizou87} showed that all
systems with a distribution function not bounded in the limit of $L\to 0$ are unstable to ROI. Such
a feature is for example characteristic of \emph{all} distribution functions of the form given in
equation \eqref{eq_df_lpart}, for which $\beta_0=\beta(r=0)>0$. Therefore, we recommend against
using the notion of the global (or half-mass) critical value of $Q$, as the onset of
ROI seems to depend predominantly on the shape of the anisotropy parameter profile $\beta(r)$.

Final axis ratios obtained in our simulations are similar to those reported by other authors.
However, one has to bear in mind that a direct comparison of the final axis ratios is difficult,
as we measured \emph{profiles} of those ratios, whereas usually only global values are provided.
The maximum elongation found by other authors is of the order of $c/a\approx 0.3$
(\citealt{meza_zamorano97}; \citealt{nipoti02, nipoti11}).
\citet{merritt_aguilar85} describe all their haloes as having $c/a=0.4-0.5$.
Haloes which form as a result of cold collapse usually have $c/a\approx 0.5$.
Dark matter haloes from cosmological simulations are typically less elongated and have smaller
differences between the inner and outer values of $c/a$. We suppose that a gradual formation of
such a halo, in contrast with the monolithic collapse, may be one of the reasons for the
difference.

The density profile of our simulated haloes did not change significantly, thus particles did not change
their mean distance form the potential centre, but only reconfigured themselves.
The decrease of the logarithmic slope $\gamma$ in the centre may be a continuation of the effect noticed by
\citet{huss_jain_steinmetz99}, i.e. the transition from $\gamma=2$ to $\gamma=1$ in the centre of the collapsed halo if the
onset of ROI was permitted.

The anisotropy parameter $\beta$ calculated in spherical shells is definitely an inadequate quantity to describe the velocity distribution of a non-spherical object.
The fact that the orbit distribution is tangentially biased in the plane perpendicular to the major axis of the halo is connected to the prevalence of orbits with significant angular momentum (O, S and I with high initial $L_{\rm x}$) in this region.
Moreover, when members of the families B, R and the rest of the family I cross this plane, they usually have small radial and large tangential components of the velocity.
On the other hand, in the region along the major axis there is much more diversity in the behaviour of the particles, thus the $\beta$ parameter remains almost the same as initially.
Altogether, it is likely that such a complicated distribution of orbits is not a result of ROI but rather an effect of bi- or triaxiality of the potential generated by the mass distribution.

A number of authors (e.g. \citealt{meza_zamorano97}) speculated about the relation between the box orbits and ROI.
\citet*{cincotta_nunez_muzzio96} studied the influence of the $l=2$ perturbation on two-dimensional orbits.
They argued that during precession the orbit with a small generalized momentum $p_\phi$
(which is in fact the angular momentum) would be attracted by the bar and if the initial
$p_\phi$ was small enough, the orbit would be caught by the bar.
As a result, it would become a two-dimensional equivalent of a box orbit and its $p_\phi$ would oscillate.
We have observed such oscillations for all components of the angular momentum of box orbits and
for components $L_y$ and $L_z$ of inner long-axis tubes.

Analyzing the behaviour of individual particles one can try to understand the development of the instability.
If the alignment of particles in the central part of the halo breaks spherical symmetry, particles in outer part
of the halo start to move in a triaxial potential.
Particles with a small $L$ which are situated there start to move on orbits belonging to families B and I.
Contribution of these families to the potential is again non-spherical.
Therefore, particles from further away once again orbit in a triaxial potential and their trajectories are modified.

Such a reasoning explains the difference between models C2 and C3.
In the former case there was a smaller number of particles with a small $L$, thus less particles that
could have a non-spherical contribution to the density distribution.
It explains also why for simulation D the outer part of the halo remained spherical.
Some of the particles situated there passed to family O and the rest to the others.
However, there were enough particles in family O (which initially had a large $L$) to balance the influence
of families I and B.

There is no good answer to the question why in the case of models with constant $\beta$ (C2 and C3) the
shells lying further from the centre of the potential are rounder than the ones lying closer to the halo centre.
It might be related to the bending instability, whose influence depends on the radius in a non-trivial manner.
The size of the area influenced by ROI may depend not on \emph{relative} parameters, e.g. the ratio
of the pericentre distance to the size of the orbit or the ratio of the initial angular momentum to its maximum
value at a given distance, but it may depend on \emph{absolute} values.
Particles with a larger mean orbit radius do not pass close to the centre.
They also have a larger angular momentum -- even if it is small compared to $L_\rmn{max}$.
Simply put, in order to get to families B and I a small absolute value of $L$ is necessary, not relative
to $L_\rmn{max}$.

\section{Summary}
\label{sec_summary}

In this paper we examined the impact of ROI on the shape of dark matter haloes.
For this purpose we performed seven different simulations of haloes with the same NFW density profile of
virial radius $r_\rmn{v}=25.8$ kpc and concentration $c=20$.
The virial mass of each halo was equal to $M_\rmn{v}=10^9 \rmn{M}_{\sun}$ corresponding in size to
haloes of dwarf galaxies obtained in cosmological simulations.

However, each model had a different initial radial dependence of the anisotropy parameter $\beta$.
Three of them had a constant anisotropy, including an isotropic one.
The next model was rather peculiar: it had a strongly anisotropic core but an isotropic envelope.
The last three had and increasing $\beta(r)$ profile, similar to haloes obtained in cosmological simulations.
They differed in $r_\rmn{t}$ -- the transition scale between the isotropic core and the anisotropic envelope.
The evolution of each halo was followed for 10 Gyr using an $N$-body code.

Among all seven haloes, four underwent ROI: two with the constant, non-zero anisotropy (C2 and C3),
the only one with the decreasing $\beta$ (D) and
the one with an increasing $\beta(r)$ and the smallest $r_\rmn{t}$ (I1).
In the three remaining haloes we did not notice any significant evolution of the shape.
In all haloes that underwent the instability the final profiles of the axis ratios turned out to depend on radius.
With the exception of model I1, haloes were elongated in the centres and remained spherical in the outer parts.
Halo I1 was spherical in the centre and had a minimum of the $c/a$ profile at a distance of about 2 kpc
from the centre. All haloes became approximately axisymmetric and prolate.

The density profiles of the haloes changed only slightly, which means that there were no large scale flows of matter.
The profiles of $\beta$ measured in the planes perpendicular to the major axes of the haloes decreased significantly, whereas the ones measured along major axes remained almost unchanged. This indicates that the velocity distribution is axisymmetric.
The mean $L$ of the haloes increased remarkably, in particular there were few particles with $L\approx 0$.
This is due to large oscillations of the angular momentum components perpendicular to the symmetry axis.
The component of the angular momentum along the symmetry axis is conserved with great accuracy.

Depending on the initial angular momentum, particles eventually join different orbit families.
Regions of membership are well separated in the planes of the angular momentum components.
Moreover, regions of different types of contribution to the halo shape also seem to be separable.
For a large initial $L$ the particle contribution is oblate.
Particles with a small $L$ join the families which make the shape of the halo more prolate.
Additionally, for small $L_x$ (component along the later symmetry axis) orbits are not axisymmetric,
thus they contribute to the triaxiality of the halo and this may be related to the fact that the triaxiality increases
with the anisotropy of the system.
Different reactions to the bar-like perturbation arise from different numbers of particles
with a given angular momentum that may later join orbit families which make the halo non-spherical.

Considering the influence of ROI on the real dark matter haloes one has to bear in mind a few important issues.
First, the infalling matter moves on rather elongated orbits and has small angular momentum relative to
the centre of the halo.
Thus, it is quite likely that ROI sets on during the halo growth making it triaxial or prolate.
However, in reality haloes accumulate mass continuously, so the situation is different than the one
considered in this work.
Second, it is unlikely that many haloes have a radial orbit distribution in their centre --
in other words, in simulated haloes we never have $\beta(r=0)$ significantly above zero.
Our simulations apply directly to haloes of dwarf galaxy size, but it seems that simple scaling
is enough to
adapt them to larger objects. One has to remember also that these results refer to isolated haloes.
In the case of an interaction of a halo with another object some of the properties discussed above may change.

\section*{Acknowledgements}

This research was partially supported by the Polish National Science Centre under grant 2013/10/A/ST9/00023.
GG acknowledges the summer student program of the Copernicus Center in Warsaw.
GG and EL{\L} are grateful for the hospitality of the Dark Cosmology Centre in Copenhagen during
their visit. The Dark Cosmology Centre is funded by the Danish National Research Foundation. We thank A. Pollo for valuable comments.

\end{document}